\documentstyle[twoside,epsf]{article}

\catcode`\@=11
\long\def\@makefntext#1{
\protect\noindent \hbox to 3.2pt {\hskip-.9pt  
$^{{\eightrm\@thefnmark}}$\hfil}#1\hfill}		

\def\thefootnote{\fnsymbol{footnote}}
\def\@makefnmark{\hbox to 0pt{$^{\@thefnmark}$\hss}}	
	
\def\ps@myheadings{\let\@mkboth\@gobbletwo
\def\@oddhead{\hbox{}
\rightmark\hfil\eightrm\thepage}   
\def\@oddfoot{}\def\@evenhead{\eightrm\thepage\hfil
\leftmark\hbox{}}\def\@evenfoot{}
\def\sectionmark##1{}\def\subsectionmark##1{}}

\oddsidemargin=\evensidemargin
\renewcommand{\thefootnote}{\fnsymbol{footnote}}

\newcounter{sectionc}\newcounter{subsectionc}\newcounter{subsubsectionc}
\renewcommand{\section}[1] {\vspace{12pt}\addtocounter{sectionc}{1} 
\setcounter{subsectionc}{0}\setcounter{subsubsectionc}{0}\noindent 
	{\tenbf\thesectionc. #1}\par\vspace{5pt}}
\renewcommand{\subsection}[1] {\vspace{12pt}\addtocounter{subsectionc}{1} 
	\setcounter{subsubsectionc}{0}\noindent 
	{\bf\thesectionc.\thesubsectionc. {\kern1pt \bfit #1}}\par\vspace{5pt}}
\renewcommand{\subsubsection}[1] {\vspace{12pt}\addtocounter{subsubsectionc}{1}
	\noindent{\tenrm\thesectionc.\thesubsectionc.\thesubsubsectionc.
	{\kern1pt \tenit #1}}\par\vspace{5pt}}
\newcommand{\nonumsection}[1] {\vspace{12pt}\noindent{\tenbf #1}
	\par\vspace{5pt}}

\newcounter{appendixc}
\newcounter{subappendixc}[appendixc]
\newcounter{subsubappendixc}[subappendixc]
\renewcommand{\thesubappendixc}{\Alph{appendixc}.\arabic{subappendixc}}
\renewcommand{\thesubsubappendixc}
	{\Alph{appendixc}.\arabic{subappendixc}.\arabic{subsubappendixc}}

\renewcommand{\appendix}[1] {\vspace{12pt}
        \refstepcounter{appendixc}
        \setcounter{figure}{0}
        \setcounter{table}{0}
        \setcounter{lemma}{0}
        \setcounter{theorem}{0}
        \setcounter{corollary}{0}
        \setcounter{definition}{0}
        \setcounter{equation}{0}
        \renewcommand{\thefigure}{\Alph{appendixc}.\arabic{figure}}
        \renewcommand{\thetable}{\Alph{appendixc}.\arabic{table}}
        \renewcommand{\theappendixc}{\Alph{appendixc}}
        \renewcommand{\thelemma}{\Alph{appendixc}.\arabic{lemma}}
        \renewcommand{\thetheorem}{\Alph{appendixc}.\arabic{theorem}}
        \renewcommand{\thedefinition}{\Alph{appendixc}.\arabic{definition}}
        \renewcommand{\thecorollary}{\Alph{appendixc}.\arabic{corollary}}
        \renewcommand{\theequation}{\Alph{appendixc}.\arabic{equation}}
        \noindent{\tenbf Appendix \theappendixc #1}\par\vspace{5pt}}
\newcommand{\subappendix}[1] {\vspace{12pt}
        \refstepcounter{subappendixc}
        \noindent{\bf Appendix \thesubappendixc. {\kern1pt \bfit #1}}
	\par\vspace{5pt}}
\newcommand{\subsubappendix}[1] {\vspace{12pt}
        \refstepcounter{subsubappendixc}
        \noindent{\rm Appendix \thesubsubappendixc. {\kern1pt \tenit #1}}
	\par\vspace{5pt}}

\newcommand{\textlineskip}{\baselineskip=13pt}
\newcommand{\smalllineskip}{\baselineskip=10pt}

\def\eightcirc{
\begin{picture}(0,0)
\put(4.4,1.8){\circle{6.5}}
\end{picture}}
\def\eightcopyright{\eightcirc\kern2.7pt\hbox{\eightrm c}} 

\newcommand{\copyrightheading}[1]
	{\vspace*{-2.5cm}\smalllineskip{\flushleft
	{\footnotesize International Journal of Modern Physics A, #1}\\
	{\footnotesize $\eightcopyright$\, World Scientific Publishing
	 Company}\\
	 }}

\def\abstracts#1#2#3{{
	\centering{\begin{minipage}{4.5in}\baselineskip=10pt\footnotesize
	\parindent=0pt #1\par 
	\parindent=15pt #2\par
	\parindent=15pt #3
	\end{minipage}}\par}} 

\renewenvironment{thebibliography}[1]
	{\frenchspacing
	 \ninerm\baselineskip=11pt
	 \begin{list}{\arabic{enumi}.}
	{\usecounter{enumi}\setlength{\parsep}{0pt}
	 \setlength{\leftmargin 12.7pt}{\rightmargin 0pt} 
	 \setlength{\itemsep}{0pt} \settowidth
	{\labelwidth}{#1.}\sloppy}}{\end{list}}

\newcounter{itemlistc}
\newcounter{romanlistc}
\newcounter{alphlistc}
\newcounter{arabiclistc}

\newcommand{\fcaption}[1]{
        \refstepcounter{figure}
        \setbox\@tempboxa = \hbox{\footnotesize Fig.~\thefigure. #1}
        \ifdim \wd\@tempboxa > 5in
           {\begin{center}
        \parbox{5in}{\footnotesize\smalllineskip Fig.~\thefigure. #1}
            \end{center}}
        \else
             {\begin{center}
             {\footnotesize Fig.~\thefigure. #1}
              \end{center}}
        \fi}

\newcommand{\tcaption}[1]{
        \refstepcounter{table}
        \setbox\@tempboxa = \hbox{\footnotesize Table~\thetable. #1}
        \ifdim \wd\@tempboxa > 5in
           {\begin{center}
        \parbox{5in}{\footnotesize\smalllineskip Table~\thetable. #1}
            \end{center}}
        \else
             {\begin{center}
             {\footnotesize Table~\thetable. #1}
              \end{center}}
        \fi}

\def\@citex[#1]#2{\if@filesw\immediate\write\@auxout
	{\string\citation{#2}}\fi
\def\@citea{}\@cite{\@for\@citeb:=#2\do
	{\@citea\def\@citea{,}\@ifundefined
	{b@\@citeb}{{\bf ?}\@warning
	{Citation `\@citeb' on page \thepage \space undefined}}
	{\csname b@\@citeb\endcsname}}}{#1}}

\newif\if@cghi
\def\cite{\@cghitrue\@ifnextchar [{\@tempswatrue
	\@citex}{\@tempswafalse\@citex[]}}
\def\citelow{\@cghifalse\@ifnextchar [{\@tempswatrue
	\@citex}{\@tempswafalse\@citex[]}}
\def\@cite#1#2{{$\null^{#1}$\if@tempswa\typeout
	{IJCGA warning: optional citation argument 
	ignored: `#2'} \fi}}

\def\pmb#1{\setbox0=\hbox{#1}
	\kern-.025em\copy0\kern-\wd0
	\kern.05em\copy0\kern-\wd0
	\kern-.025em\raise.0433em\box0}

\def\fnt#1#2{\footnotetext{\kern-.3em
	{$^{\mbox{\scriptsize #1}}$}{#2}}}

\def\fpage#1{\begingroup
\voffset=.3in
\thispagestyle{empty}\begin{table}[b]\centerline{\footnotesize #1}
	\end{table}\endgroup}

\def\runninghead#1#2{\pagestyle{myheadings}
\markboth{{\protect\footnotesize\it{\quad #1}}\hfill}
{\hfill{\protect\footnotesize\it{#2\quad}}}}
\font\tenrm=cmr10
\font\tenit=cmti10 
\font\tenbf=cmbx10
\font\bfit=cmbxti10 at 10pt
\font\ninerm=cmr9

\font\eightrm=cmr8

\def\qed{\hbox{${\vcenter{\vbox{			
   \hrule height 0.4pt\hbox{\vrule width 0.4pt height 6pt
   \kern5pt\vrule width 0.4pt}\hrule height 0.4pt}}}$}}

\renewcommand{\thefootnote}{\fnsymbol{footnote}}	

\begin{document}

\runninghead{Determining Color-Octet Matrix Elements $\ldots$} 
            {Determining Color-Octet Matrix Elements $\ldots$}

\normalsize\textlineskip
\thispagestyle{empty}
\setcounter{page}{1}

\copyrightheading{}			

\vspace*{0.88truein}

\fpage{1}
\centerline{ \bf Determining Color-Octet $\psi$-Production Matrix Elements}
\vspace*{0.035truein}
\centerline{ \bf from $\gamma p$ and $ep$ Processes\footnote{Presented
at the 1996 University of Illinois at Chicago Quarkonium Physics
Workshop. This work was done in collaboration with Jim Amundson, Ivan
Maksymyk, and Tom Mehen.}} 
\vspace*{0.37truein}
\centerline{ \footnotesize Sean Fleming}
\vspace*{0.015truein}
\centerline{ \footnotesize\it Department of Physics, University of Wisconsin}
\vspace*{0.225truein}

\vspace*{0.21truein}
\abstracts{
We calculated, within the NRQCD factorization
formalism, the leading color-octet contributions to $\psi$ production
in photon-nucleon and electron-nucleon collisions. The expressions
obtained depend on the NRQCD matrix elements 
$\langle {\cal O}^{\psi}_8(^1S_0) \rangle$ and 
$\langle {\cal O}^{\psi}_8(^3P_J) \rangle$. These matrix elements can
be determined by fitting to experimental data. The color-octet
contribution to $\psi$ photoproduction is in the
forward region of phase space, where there may be large
corrections to the NRQCD result from higher twist terms.
As to $\psi$ leptoproduction
we point out that the theoretical uncertainties plaguing the
photoproduction calculation vanish in the large momentum transfer
limit. In this region of phase space the NRQCD formalism should be
valid, making $\psi$ leptoproduction   
an ideal laboratory for testing the theory.}{}{}


\vspace*{1pt}\textlineskip	
\section{Introduction}		
\vspace*{-0.5pt}
\noindent
The nonrelativistic QCD (NRQCD) factorization formalism developed by
Bodwin, Braaten, and Lepage provides a rigorous theoretical framework
within which quarkonium production can be studied~\cite{bbl}. A
central result of this formalism is that 
the cross section for the inclusive production of a
quarkonium state $H$ is a sum of products having the form
\begin{equation}
\sigma (A+B \to H+X) = \sum_n {F_n \over m^{d_n -4}_Q} 
\langle {\cal O}^{H}_n \rangle \; ,
\label{ffcs}
\end{equation}
where $m_Q$ is the mass of the heavy quark $Q$, $F_n$ are
short-distance coefficients, and ${\cal O}^H_n$ are NRQCD four-fermion
production operators with naive energy dimension $d_n$. The short-distance
coefficients, $F_n$, are associated with the production of 
a $Q\bar{Q}$ pair with quantum numbers indexed by $n$ (angular
momentum ${}^{2S+1}L_J$ and color 1 or 8). They can be calculated using
perturbative techniques. The NRQCD production matrix elements,
{}$\langle{\cal O}^H_n\rangle\equiv\langle 0|{\cal O}^H_n|0\rangle$, 
parameterize the hadronization into $H$ of a $Q\bar{Q}$ pair with
quantum numbers indexed by $n$. They can be determined phenomenologically.
\pagebreak

\textheight=7.8truein
\setcounter{footnote}{0}
\renewcommand{\thefootnote}{\alph{footnote}}

The power of the NRQCD formalism stems from the fact that
Eq.~(\ref{ffcs}) is essentially an expansion in the small parameter
{}$v^2$, where $v$ is the average relative velocity of the $Q$ and
{}$\bar{Q}$ in the boundstate $H$. $v^2 \sim 0.3$ for
charmonium. NRQCD $v$-scaling rules~\cite{lmnm} allow one to estimate
the relative 
sizes of the various $\langle{\cal O}^H_n\rangle$. This information,
along with knowlege of the dependence of the $F_n$ on coupling
constants, permits one to decide which terms must be retained in
expressions for observables to reach a a given level of
accuracy. Generally, to leading order, factorization formulas involve
only a few matrix elements, so several observables can be related by a
small number of parameters. Thus it is possible to test the NRQCD
factorization formalism by determining if a body of data can be
consistently fit by the most important NRQCD production matrix elements.
Moreover,
by studying the regime in which the NRQCD factorization formalism
fails we can gain valuable insight into the limitations of the theory. 

As to $J/\psi$ production, in many instances the most important NRQCD
matrix elements are  
{}$\langle{\cal O}^{\psi}_1(^3S_1)\rangle$,
{}$\langle{\cal O}^{\psi}_8(^3S_1)\rangle$,
{}$\langle{\cal O}^{\psi}_8(^1S_0)\rangle$, and 
{}$\langle{\cal O}^{\psi}_8(^3P_J)\rangle$. These matrix elements
appear in the expressions of rates for $\psi$ production in hadronic
collisions, in $Z^0$ decay, in $e^+ e^-$ annihilation, in
photon-nucleon collisions, and in lepton-nucleon
collisions~\cite{review}. Fitting the leading order predictions to 
the various experimental data has revealed some
inconsistencies~\cite{photopro}~\cite{br1}. In particular 
the value determined for the linear combination 
{}$\langle{\cal O}^{\psi}_8(^1S_0)\rangle + 3 
{}\langle{\cal O}^{\psi}_8(^3P_0)\rangle/m^2_c$ at CDF~\cite{cl}
appears to be 
incompatible with the value determined for the linear combination 
{}$\langle{\cal O}^{\psi}_8(^1S_0)\rangle + 7
{}\langle{\cal O}^{\psi}_8(^3P_0)\rangle/m^2_c$
from photoproduction~\cite{photopro}~\cite{afm} and other 
hadroproduction~\cite{br2} experiments. Another aspect of this problem
is that if one uses the CDF measurement to make an estimate of the
magnitude of {}$\langle{\cal O}^{\psi}_8(^1S_0)\rangle$ 
and {}$\langle{\cal O}^{\psi}_8(^3P_J)\rangle$, assuming both matrix
elements to be positive, 
the color-octet contribution to inelastic $\psi$
photoproduction is too large~\cite{photopro}. Throughout the paper we
will refer to this inconsistency as the ``photoproduction conundrum''. 

In this work we focus on aspects of $\psi$ photoproduction and
electroproduction. Within the context of these processes we
investigate some of the issues involved in 
applying the NRQCD factorization formalism, and
study the possible limitations of the theory. More precisely our goal
is to see if an analysis of $\psi$ leptoproduction can resolve the
photoproduction conundrum.

The paper is divided into two parts. The first part, section 2, is a
calculation of the forward $\psi$ photoproduction rate. We fit the
theoretical expression for the rate to experimental data and 
determine a value for the linear combination 
{}$\langle{\cal O}^{\psi}_8(^1S_0)\rangle + 
7 \langle{\cal O}^{\psi}_8(^3P_J)\rangle/m^2_c$. We point out that
there are large corrections to this result; a sign that 
this process is not amenable to analysis within the NRQCD formalism.
 
In the second part of the paper, section 3, we present a calculation
of the $\psi$ 
leptoproduction rate. For large momentum-transfer squared, $Q^2$, this
process does not suffer the sizable corrections afflicting the
photoproduction calculation, and is, therefore,  well within the
regime of applicability of the NRQCD formalism. Unfortunately, at this
time, there is no data 
available on forward $\psi$ leptoproduction at large $Q^2$. Thus
we fit the theoretical result to experimental data covering low to
moderate values of $Q^2$, and measure  
{}$\langle{\cal O}^{\psi}_8(^1S_0)\rangle$ and
{}$\langle{\cal O}^{\psi}_8(^3P_0)\rangle$. We find a negative value
for {}$\langle{\cal O}^{\psi}_8(^3P_0)\rangle$. By studying the
renormalization of the $P$-wave operator we argue that this is not
shocking. The values of the color-octet matrix elements determined
here resolve one aspect of the photoproduction conundrum: the CDF
analysis is now consistent with photoproduction and other
hadroproduction analyses. It is unclear, however, whether 
the other aspect of the photoproduction conundrum is resolved: that the
color-octet contribution to inelastic $\psi$ photoproduction is too
large.  

\section{Photoproduction of $\psi$}
\noindent
One of the earliest calculations of $\psi$ production was carried out
by Berger and Jones in 1981~\cite{bj}. In this paper the authors
present a calculation of the rate for $\psi$ production in
$\gamma$-nucleon collisions carried out within the 
color-singlet model. 
In this model one assumes that there is a nonzero
probability for a $Q\bar{Q}$ pair to form a hadron $H$ only if the
$Q\bar{Q}$ pair is produced at short-distances ({\it i.e.} at
distances on the order of $1/m_Q$ or less) with the quantum numbers of
the dominant Fock state of $H$~\cite{schuler}. For example a
{}$c\bar{c}$ pair has a nonzero probability of forming a $\psi$ only
if the $c\bar{c}$ pair is produced at short-distances in a
color-singlet state with angular-momentum quantum numbers ${}^3S_1$.

In order to gain a deeper understanding of the color-singlet results 
we need to define the parameter
{}$z \equiv N\cdot P/N\cdot k$. Here $N$ is the initial-state nucleon 
four-momentum, $k$ is the initial-state photon four-momentum, and
{}$P$ is the $\psi$ four-momentum. In the rest frame of the
nucleon $z$ is the 
fraction of the photon energy that is carried away by the $\psi$. The
region of phase space where  
{}$z < 0.9$ is, by convention, defined as ``inelastic''. The remaining
region of phase space $0.9 < z < 1.0$ is defined as ``forward''.

If one includes next-to-leading order QCD corrections and
next-to-leading order relativistic corrections the
color-singlet model explains the experimental data in the
inelastic region~\cite{nlophoto}. However, the color-singlet model
prediction falls nearly an order of magnitude below the data in the
forward region~\cite{photopro}. This is not unexpected since, as
Berger and Jones pointed out over a decade ago~\cite{bj}, in the
forward region of phase space the final state gluon couples to
on-shell quark lines; 
thus {}$\alpha_s$ for the vertex is ill-defined and the emission of
the gluon is nonperturbative. 

The failure of the color-singlet model to explain forward $\psi$
photoproduction leads us to the question: can $\psi$
photoproduction be understood in the NRQCD factorization formalism?

According to the NRQCD factorization formalism the photoproduction
cross section is given by Eq.~\ref{ffcs}. The $v$-scaling rules tell
us that the probability for producing 
$H$ from a $Q\bar{Q}$ pair that does not have the quantum numbers of
the dominant Fock state of $H$ is down by powers of $v^2$
relative to the probability for producing $H$ from a $Q\bar{Q}$ pair
with the quantum numbers of the dominant Fock state of $H$. As pointed out
previously, $v^2$ is not in general zero so it is possible for a
{}$Q\bar{Q}$ pair  produced at short-distances with any quantum numbers
to hadronize into $H$. Only in the $v \to 0$ limit is the
color-singlet model recovered.

Within the NRQCD factorization formalism the
leading contributions to the photoproduction cross section come
from the production of a $c\bar{c}$ pair in a color-singlet ${}^3S_1$
state, or from the production of a $c\bar{c}$ pair in a color-octet
state in either a ${}^1S_0$ or ${}^3P_J$ 
configuration. To leading order in $\alpha_s$ the color-singlet
short-distance coefficient $F_1(^3S_1)$ can be determined 
from the diagrams in figure~1. The result is
proportional to the expression derived for the $\psi$ photoproduction
cross section in the color-singlet model~\cite{bj}. 
\begin{figure}
\begin{center}
\epsfxsize=0.4\hsize
\mbox{\epsffile{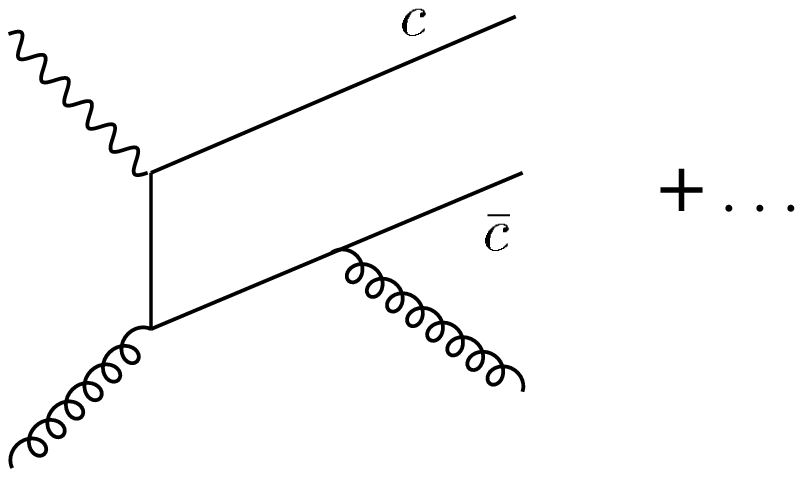}}
\end{center}
\fcaption{Leading order diagrams for the photoproduction of a 
{}$c\bar{c}$ in a color-singlet, ${}^3S_1$ state.}
\end{figure}
To leading order in $\alpha_s$ the color-octet short-distance
coefficients $F_8(^1S_0)$ and $F_8(^3P_J)$ can be
determined from the Feynman diagrams in figure~2.
\begin{figure}
\begin{center}
\epsfxsize=0.5\hsize
\mbox{\epsffile{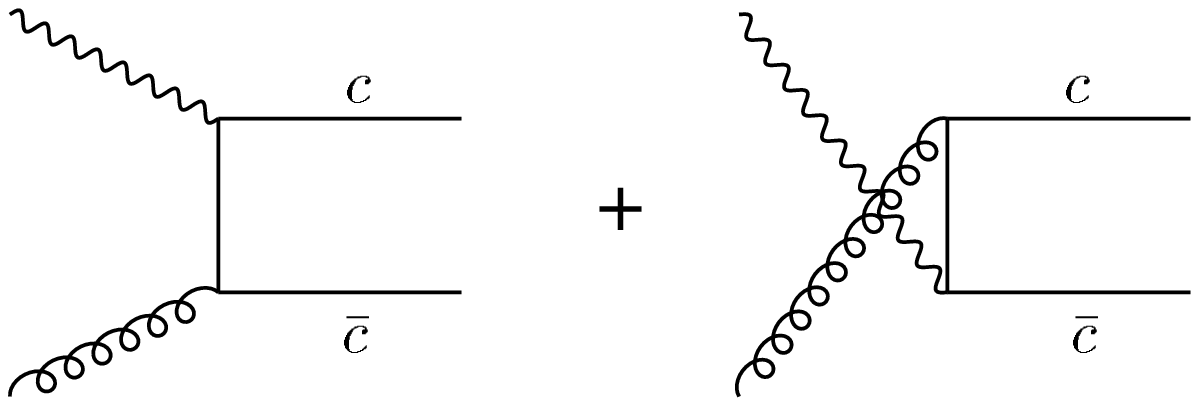}}
\end{center}
\fcaption{Leading order diagrams for the photoproduction of a 
{}$c\bar{c}$ in a color-octet state with angular momentum
configuration ${}^1S_0$, ${}^3P_0$, and ${}^3P_2$.}
\end{figure}
The color-octet matrix elements are suppressed by $v^4 \approx 0.1$
relative to 
the color-singlet matrix element, but the color-octet short-distance
coefficients are enhanced by a $\pi /\alpha_s(2m_c) \approx 10$
relative to the color-singlet short-distance coefficient. Thus both
contributions are equally important.

The NRQCD factorization formalism separates effects of short-distance
scales of order {}$m_c$ or higher, which are associated with the 
production of the 
{}$c\bar{c}$, from long-distance scales such as $m_c v$, which are
associated with the hadronization of the $c \bar{c}$ into the $\psi$.
This separation is embodied in the factored form of the production
cross section as shown in Eq.~(\ref{ffcs}). 
The spirit of the formalism is that 
{}$\gamma + g \to c\bar{c}_1(^3S_1) + g$ is the correct short-distance
process as long as the emission of the final state gluon takes place
within a distance of order $1/m_c$ of the interaction point. This
means that this gluon must have momentum of order $m_c$ or
greater. If the final state gluon has momentum less than $m_c$ its
emission is not part of the short-distance process and the diagrams
shown in figure~1 do not describe the hard scattering. 

Therefore, we hypothesize that the color-singlet calculation is not
valid for the region of phase space $0.9 < z < 1.0$, since in
this region the emission of the final state gluon is nonperturbative,
and is thus not part of the short distance process. Rather
photoproduction in the forward region is described by the leading
color-octet process. Inspired by the NRQCD formalism we will limit the
color-singlet contribution to the region of 
phase space where $0< z < 1- \Lambda$, where $\Lambda$ is some
arbitrary cutoff of order $v^2$.
Then the color-octet contribution produces $\psi$ in the region 
{}$1-\Lambda < z < 1$. By convention in experiments the cutoff point
between inelastic and forward $\psi$ production is chosen to be 
{}$z \approx 0.9$. Thus we choose $\Lambda \approx 0.1$. We wish to
emphasize the cutoff does not arise naturally in the NRQCD formalism,
rather it is an assumption made in the spirit of the formalism.

The color-singlet contribution to $\psi$ photoproduction has been
studied extensively~\cite{nlophoto}. Let us now consider the
color-octet contribution. 

The leading color-octet contribution to the
$\psi$ photoproduction cross section can be calculated from the Feynman
diagrams given in figure~2. The resulting expression is~\cite{afm}

\begin{equation}
\sigma(\gamma + N \to \psi + X) = \int dx \; f_{g/N}(x) \;
{\alpha_s(2m_c) \alpha e^2_c \pi^3 \over m^3_c} \;
\delta(xs-4m^2_c) \; \Theta \; ,
\label{cocs}
\end{equation}
where $x$ is the momentum fraction of the incoming gluon relative to
the nucleon, $f_{g/N}(x)$ is the gluon distribution function for the
nucleon, and 
\begin{equation}
\Theta = \langle{\cal O}^{\psi}_8(^1S_0)\rangle + 7
{\langle{\cal O}^{\psi}_8(^3P_0)\rangle \over m^2_c} \; .
\label{theta}
\end{equation}
Since the values of the color-octet matrix elements are not known we 
can not make a prediction for the forward photoproduction cross section. 
However, we can fit the results to experimental data and make a 
prediction for $\psi$ production in some other process. 

Figure~3 shows a fit of Eq.~(\ref{cocs}) to forward cross-section
measurements from the fixed target experiments E687~\cite{e687},
NA14~\cite{na14}, E401~\cite{e401}, NMC~\cite{nmc}, 
and E516~\cite{e516}. 
\begin{figure}
\begin{center}
\epsfxsize=0.5\hsize
\mbox{\epsffile{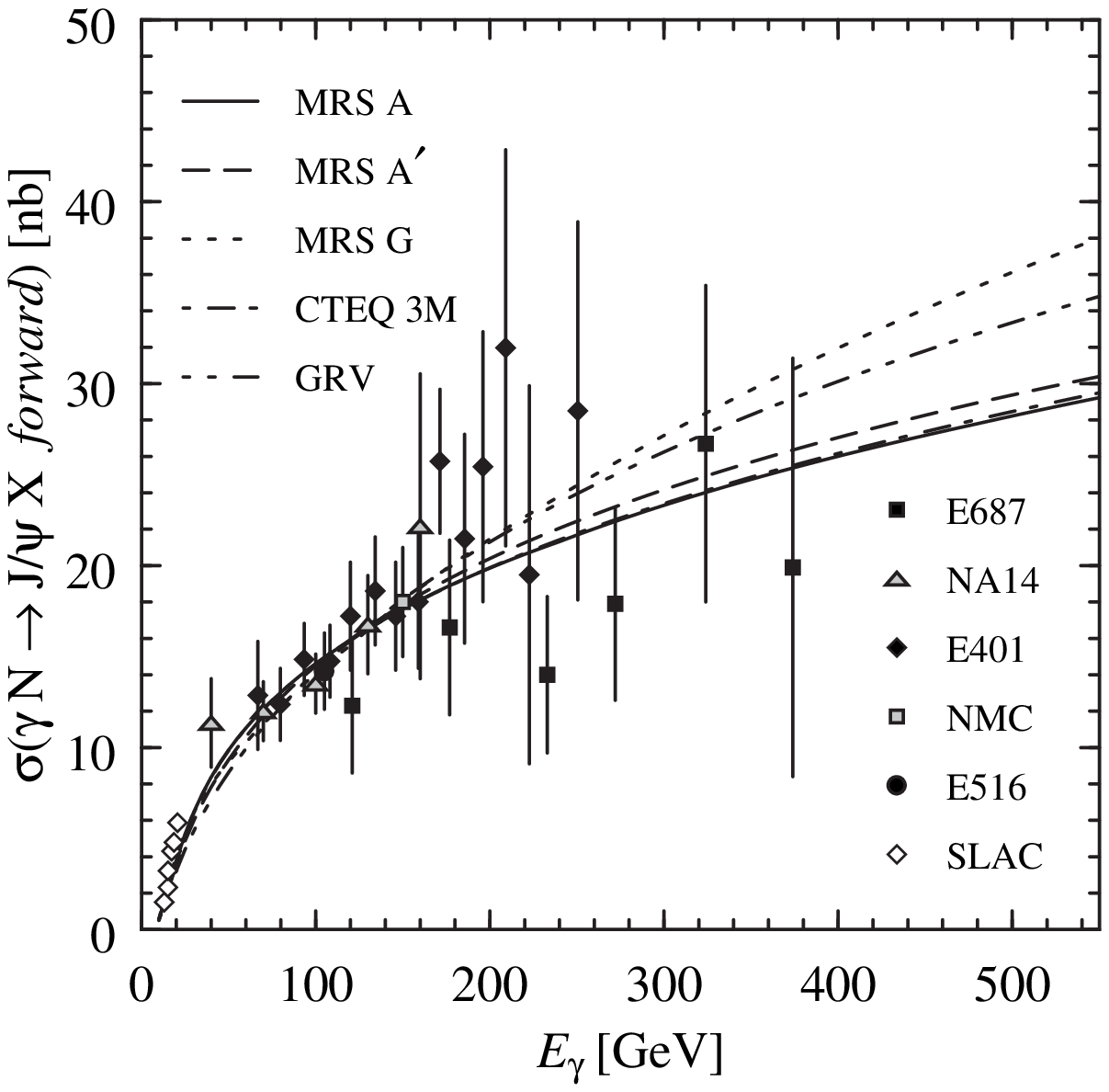}}
\end{center}
\fcaption{A one parameter fit to experimental data of the NRQCD factorization 
formalism result for forward photoproduction of $\psi$.}
\end{figure}
Using $\alpha_s(2m_c) = 0.26$ and $m_c = 1.5$~GeV we obtain the value
\begin{equation}
\label{Thetaw}
\Theta = 0.02 \mbox{~GeV}^{3} \; ,
\end{equation}
No theoretical error has been quoted here. The expression given in 
Eq.~(\ref{cocs}) is very sensitive to the value we choose for the parameter 
$m_{c}$, which results in a large uncertainty in the value determined 
for $\Theta$. Given this error, the number presented in Eq.~(\ref{Thetaw}) 
is consistent with the value {}$\Theta = 0.03 \mbox{~GeV}^{3}$ measured 
in $\pi N$ collisions~\cite{br2}.

There is an important point to make regarding the calculation of
forward $\psi$ photoproduction. Namely, in the derivation of the
factorization formula presented in Eq~(\ref{ffcs}) higher twist terms
have been neglected. In the forward region these higher twist terms 
can be large~\cite{highertwist}. For example, they could describe
correlations between the initial and final state resulting in
diffractive and elastic processes. Such processes can 
have large contributions to the $\psi$ forward photoproduction cross 
section, and can, therefore, significantly affect the value determined 
for $\Theta$~\cite{risk}.

Given the large theoretical uncertainty associated with the forward 
photoproduction calculation it is best to regard the value presented 
in Eq.~(\ref{Thetaw}) as an upper limit. In fact, if the value 
determined for $\Theta$ in photoproduction is consistent with fits of the 
color-octet matrix elements to other $\psi$ production data we can be 
certain that higher twist corrections, and diffractive and elastic 
contributions are small.

\section{Leptoproduction of $\psi$}
\noindent
Many of the theoretical uncertainties plaguing the photoproduction 
calculation can be avoided by requiring the incoming photon to be 
highly virtual. This introduces a new scale into the process: $Q^{2}$ 
the momentum transfered through the photon squared. Higher twist 
terms will be suppressed by powers of $Q^{2}$, and will, therefore, 
vanish in the large $Q^2$ limit.

A process in which the photon can be highly virtual is $\psi$ 
leptoproduction. 
The calculation of $\psi$ leptoproduction is analogous to the
photoproduction calculation, except the incoming photon is off
shell. In the region of phase space where $0.9 < z^* < 1.0$ (the $*$
denotes that the photon is virtual) the leading
contribution comes from the fusion of the virtual photon and the gluon
into a color-octet $c\bar{c}$ pair in either a ${}^1S_0$, ${}^3P_0$,
or ${}^3P_2$ configuration, followed by the hadronization of the
{}$c\bar{c}$ pair into a $\psi$. The Feynman diagrams used to
determine the short-distance coefficients are shown in figure~4.
\begin{figure}
\begin{center}
\epsfxsize=0.5\hsize
\mbox{\epsffile{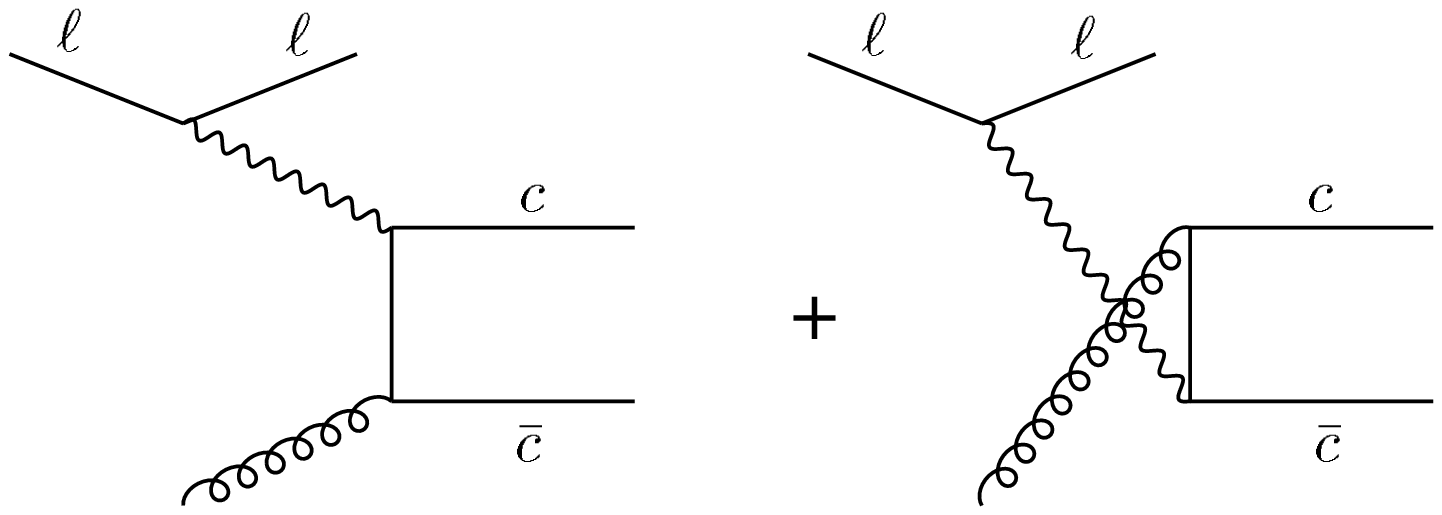}}
\end{center}
\fcaption{A two parameter fit to EMC data of the NRQCD factorization 
formalism result for forward leptoproduction of $\psi$.}
\end{figure}

The expression for the cross section determined from the diagrams in
figure~4 is~\cite{fmm}
\begin{eqnarray}
\lefteqn{
\sigma(e+p\to e+\psi +X) = \int {dQ^2 \over Q^2}\int {dy \over y}  
\int dx\; f_{g/N}(x)\; \delta(xys-4m^2_c-Q^2) }
\nonumber \\
& & \times {2 \alpha_s(\mu^2) \alpha^2 e^2_c \pi^2 \over xsm_c}
\Bigg\{ {1+(1-y)^2 \over y} 
\Big[ \langle{\cal O}^{\psi}_8(^1S_0)\rangle + 
{3Q^2+7(2m_c)^2 \over xys} 
{\langle{\cal O}^{\psi}_8(^3P_J)\rangle \over m^2_c} \Big]
\nonumber \\
& & \;\;\;\;\;\;\;\;\; -{8(2m_c)^2Q^2 \over x^2 y s^2}
{\langle{\cal O}^{\psi}_8(^3P_J)\rangle \over m^2_c} \Bigg\}
\label{epcs}
\end{eqnarray}
where $\mu^2 = Q^2 + m^2_c$, and $x$ and $f_{g/N}(x)$ are the same as
in Eq.~(\ref{cocs}). The
momentum fraction of the virtual photon relative to the incoming
lepton is {}$y \sim N\cdot q / N \cdot k$, where $N$ is the nucleon
four-momentum, $q$ is the photon four momentum, and $k$ is the incoming
lepton four-momentum. 

The result presented in Eq~(\ref{epcs}) holds 
for all values of $Q^{2}$. Taking the limit $Q^2 \to 0$ one recovers
the photoproduction result convoluted with the electron splitting function:
\begin{equation}
\lim_{Q^2\to 0}\sigma(e+P\to e+\psi+X)\to {\alpha \over 2 \pi}
\int{dQ^2 \over Q^2} \; \int^1_0 dy \; {1+(1-y)^2\over y}
\hat{\sigma}(\gamma P \to \psi) \; .
\label{Qgoto0}
\end{equation}
As discussed previously, in this limit corrections 
to the cross section from higher twist terms, may be large.
However, in the high-energy limit $Q^2, s \gg (2m_c)^2$ we expect 
contributions from higher twist terms to vanish. Letting $Q^2, s \gg
(2m_c)^2$ we obtain 
\begin{eqnarray}
\lefteqn{
\lim_{m^2_c/Q^2,m^2_c/s\to 0}\sigma(e+P\to e+ \psi+X)\to {\alpha \over 2 \pi}
\int{dQ^2 \over Q^2} \; \int dy \; {1+(1-y)^2\over y} }
\nonumber \\ 
& & \int dx \; f_{g/N}(x) {4 \alpha_s(Q^2)\alpha e^2_c \pi^3 \over Q^2}
\left( \langle{\cal O}^{\psi}_8(^1S_0)\rangle + 3
{\langle{\cal O}^{\psi}_8(^3P_J)\rangle \over m^2_c} \right) 
\; \delta(xys-Q^2) \; .
\label{helim}
\end{eqnarray}
Note that this expression does not depend very strongly on the value
chosen for $m_{c}$. Since theoretical corrections to  
Eq.~(\ref{helim}) are expected to be small, high
energy leptoproduction provides an excellent means to
measure the linear combination 
$\langle{\cal O}^{\psi}_8(^1S_0)\rangle 
+ 3 \langle{\cal O}^{\psi}_8(^3P_J)\rangle /m^2_c$. 
This is precisely the linear combination of NRQCD matrix elements 
determined from CDF data on $\psi$ production at high transverse 
momentum. Therefore, $\psi$ leptoproduction data taken in the high 
energy limit will provide us with the opportunity to test the NRQCD 
factorization formalism by measuring the same linear combination of 
matrix elements in two different processes.

As of yet there is no $\psi$ leptoproduction data that truely falls 
in the high energy regime. Therefore we will naively fit 
Eq.~(\ref{epcs}) to the available data. The result of a fit
to EMC data~\cite{emc} over the entire range of $Q^2$
($0<Q^2<15\;\mbox{Gev}^2$) is shown in
figure~5. 
\begin{figure}
\begin{center}
\epsfxsize=0.5\hsize
\mbox{\epsffile{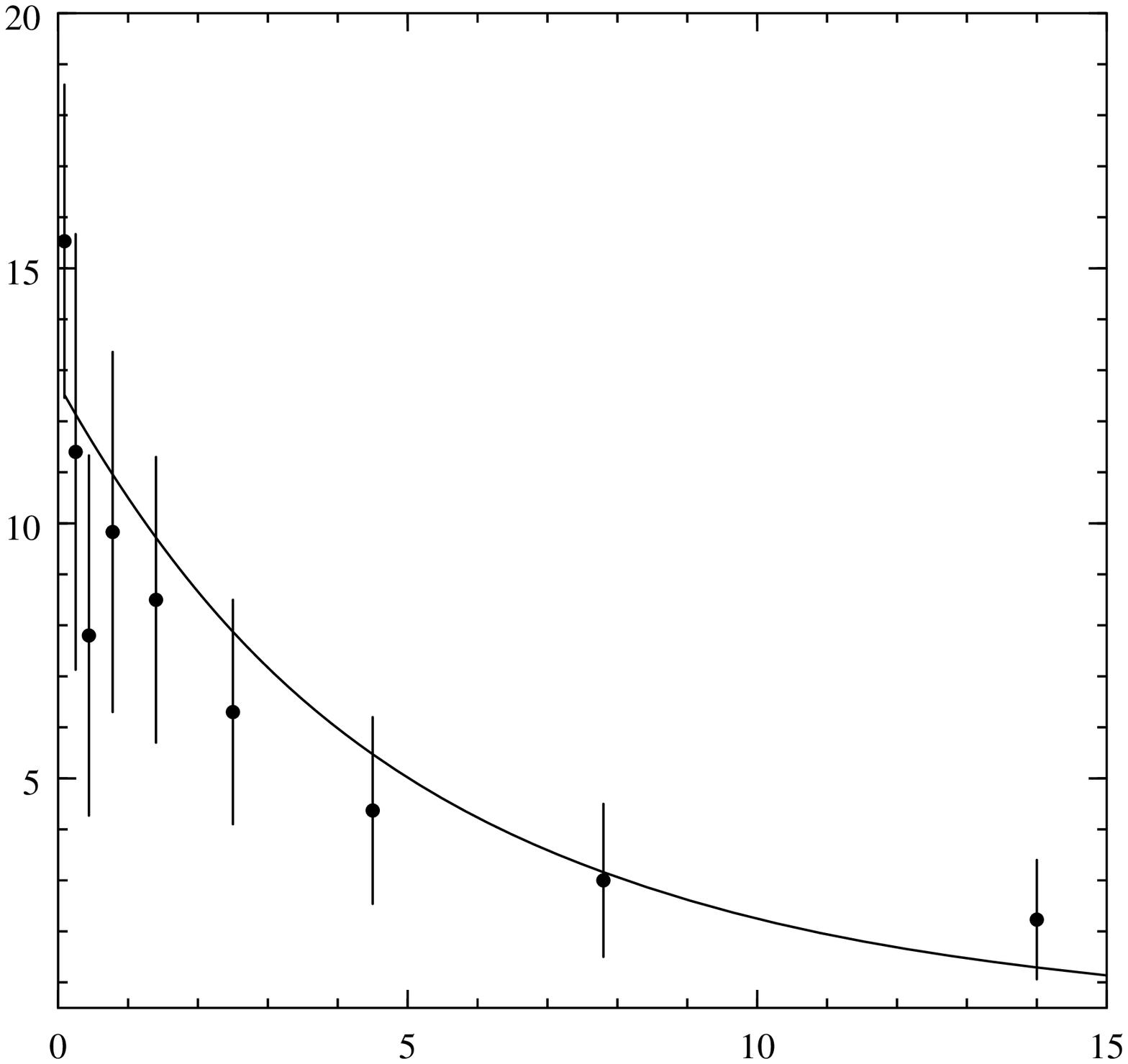}}
\end{center}
\fcaption{A two parameter fit to EMC data of the NRQCD factorization 
formalism result for forward leptoproduction of $\psi$. The horizontal
axis is $Q^2$, and the vertical axis 
is $d \sigma(\gamma^* + P \to \psi +X)/d \log{Q^2}$.}
\end{figure}
The values of the color-octet matrix elements determined
from this fit are:
\begin{eqnarray}
\langle{\cal O}^{\psi}_8(^1S_0)\rangle  & = & 0.04 \; \mbox{GeV}^3 
\; ,
\nonumber \\
{\langle{\cal O}^{\psi}_8(^3P_J)\rangle \over m^2_c} & = & -0.003
\; \mbox{GeV}^3 \; .
\label{fr}
\end{eqnarray}
Note that {}$\langle{\cal O}^{\psi}_8(^3P_J)\rangle$ is negative! Does 
this mean 
that the NRQCD factorization formalism fails to describe $\psi$
leptoproduction? Perhaps; remember there could be large
corrections to the low $Q^2$ region. However, there is another
explanation. 

Loop corrections to NRQCD operators give rise to ultraviolet power
divergences which have the form of renormalization of
lower-dimensional operators. Since dimension six is the lowest
possible energy dimension of NRQCD production operators the 
dimension-six operator 
{}${\cal O}^{\psi}_8(^1S_0)$ will not have a power-divergent
contribution. However, the dimension-eight operator 
{}${\cal O}^{\psi}_8(^3P_J)$, will have power divergences proportional 
to dimension six operators. At order $\alpha_s$, using a momentum cutoff 
$\Lambda$, these 
divergences can be removed by defining the renormalized operator to be 
\begin{equation}
{\cal O}^{\psi}_8(^3P_J)_{\mbox{r}} = 
{\cal O}^{\psi}_8(^3P_J)_{0} -
c_1 \alpha_s \Lambda^2 {\cal O}^{\psi}_1(^3S_1) -
c_8 \alpha_s \Lambda^2 {\cal O}^{\psi}_8(^3S_1) \; ,
\label{div}
\end{equation}
where the coefficients $c_{1}$ and $c_{8}$ are adjusted to cancel 
quadratic divergences at order $\alpha_{s}$. Matrix elements of the 
bare operator ${\cal O}^{\psi}_8(^3P_J)_{0}$ must indeed be positive 
definite, but this is not necessarily true for the renormalized 
operator. If one defines the operator ${\cal O}^{\psi}_8(^3P_J)$ 
using a method like dimensional regularization which automatically 
removes power divergences, the regularization method makes the above 
subtractions implicitly. Therefore the renormalized operator need not 
have positive matrix elements~\cite{eb}.

In particular it is easy to see that the 
most important subtracted term in Eq.~(\ref{div}) is the one
proportional to the
matrix element ${\cal O}^{\psi}_1(^3S_1)$. This term is 
suppressed by a factor of $\alpha_{s}$, but it is enhanced by a 
quadratic divergence and a relative factor of $1/v^{4}$. Thus it is 
not too shocking if the resulting renormalized matrix element turns 
out to be negative. Note that renormalization respects the 
$v$-scaling rules which require the $\it magnitude$ of the matrix 
element of ${\cal O}^{\psi}_8(^3P_J)$ to scale as $v^{7}$.

Accepting the explanation proffered above let us compare the results
presented in Eq.~(\ref{fr}) to photoproduction and hadroproduction results. 
The numbers determined 
in leptoproduction are consistent with the photoproduction result
given in Eq.~(\ref{Thetaw}), and are, therefore, consistent with the result of
an analysis of $\pi N$ collisions~\cite{br2}. Using CDF data on 
$\psi$ production   
at the Tevatron Cho and Leibovich have determined~\cite{cl}
\begin{equation}
\langle{\cal O}^{\psi}_8(^1S_0)\rangle + 3
{\langle{\cal O}^{\psi}_8(^3P_J)\rangle \over m^2_c} 
= 0.066 \; \; \; \mbox{GeV}^3 \; .
\label{cl}
\end{equation}
Substituting the
values given in Eq.~(\ref{fr}) into the left-hand side of
Eq.~(\ref{cl}) we obtain $0.03 \; \; \mbox{GeV}^3$. Given the large
theoretical uncertainty associated with calculations of $\psi$
production in hadronic collisions these results are consistent. 

\section{Conclusion}
\noindent
We have calculated, within the NRQCD factorization formalism, the
leading color-octet contributions to $\psi$ photoproduction and $\psi$
leptoproduction. The expressions obtained depend on the color-octet
matrix elements  $\langle {\cal O}^{\psi}_8(^1S_0) \rangle$ and 
{}$\langle {\cal O}^{\psi}_8(^3P_0) \rangle$. The NRQCD factorization
formalism may be tested by fitting these matrix elements to
experimental data, and then making predictions for $\psi$ production
in some other process. 

The color-octet contribution to the 
photoproduction cross section is in the region of phase space where
the $\psi$ is produced in the forward direction. In this region 
higher twist terms, which were neglected in the derivation of the
factorization formula Eq.~\ref{ffcs}, may be large. 
Therefore, forward $\psi$ photoproduction is not
amenable to testing the NRQCD factorization formalism. However, we can
learn about the limitations of the theory, and the size of possible
corrections. 

The leading color-octet contribution to the $\psi$ leptoproduction
cross section does not suffer from the same problems as the
photoproduction calculation if we restrict ourselves to the high
$Q^2$ regime. In this region $\psi$ leptoproduction should provide an
ideal laboratory for testing the NRQCD factorization
formalism. However, there is at this time no experimental data for
$\psi$ production in the asymptotic regime. Therefore, keeping in mind
that there may be large corrections to the cross section in the low
$Q^2$ region, we fit our results to EMC data 
for $0<Q^2<15 \; \mbox{GeV}^2$. We 
determine $\langle{\cal O}^{\psi}_8(^1S_0)\rangle = 0.04$, 
and  $\langle{\cal O}^{\psi}_8(^3P_0)\rangle = - 0.003$. 
The negative value for the $P$-wave matrix element is acceptable since
the renormalized $P$-wave operator is a subtraction of two divergent
terms, ${\cal O}^{\psi}_8(^3P_J)_{\mbox{r}} = {\cal
O}^{\psi}_8(^3P_J)_{0} - c_1 \alpha_s \Lambda^2 {\cal
O}^{\psi}_1(^3S_1)$, where the second term on the right-hand-side can
be larger than the first term on the right-hand-side. The values
determined for the color-octet matrix elements resolve one aspect of
the photoproduction conundrum: the discrepancy between the CDF analysis
and photoproduction and other hadroproduction analysis. However it is not
clear if the other aspect of the photoproduction
conundrum is resolved: the color-octet contribution to inelastic
{}$\psi$ production is too large.

\noindent
\nonumsection{Acknowledgements}
\noindent
I would like to thank my collaborators on the projects that have 
contributed to this work. They are Jim Amundson, Ivan Maksymyk, and 
Tom Mehen. I would also like to thank Eric Braaten for sharing his 
insights on renormalization of the NRQCD matrix elements. This work 
was supported in part by the University of Wisconsin Research 
Committee with funds granted by the Wisconsin Alumni Research 
Foundation, and the U.S. Department of Energy under grant 
DE-FG02-95ER40896.

\nonumsection{References}

\end{document}